\documentstyle[sprocl]{article}

\input{psfig}

\def\be{\begin{equation}}
\def\ee{\end{equation}}
\def\bea{\begin{eqnarray}}
\def\eea{\end{eqnarray}}

\begin{document}

\title{THE COMA CLUSTER IN RELATION TO ITS ENVIRONS}

\author{Michael J. West\footnote{Also CITA Senior Visiting Fellow,
Canadian Institute for Theoretical Astrophysics, \newline University of Toronto,
60 Saint George Street,
Toronto, ON M5S 1A7, Canada}}

\address{Saint Mary's University \\ Department of Astronomy \& Physics \\ 
Halifax, Nova Scotia B3H 3C3, Canada \\
E-mail: west@sisyphus.stmarys.ca}

\maketitle\abstracts{Clusters of galaxies are often embedded in 
larger-scale superclusters with dimensions of tens or perhaps even
hundreds of Mpc.  Observational and theoretical evidence  
suggest an important connection between cluster properties
and their surroundings, with cluster formation being driven primarily
by the infall of material along large-scale filaments.  
Nowhere is this connection more obvious than the Coma cluster.   
}

\section{Historical Background}

The Coma cluster's surroundings have been discussed in the
astronomical literature for nearly as long as the cluster itself.
William Herschel (1785) discovered what he called
``the nebulous stratum of Coma Berenices'' 
and remarked that ``I have fully ascertained
the existence and direction of this stratum for more than 30 degrees
of a great circle and found it to be almost every where equally
rich in fine nebulae.''
Coma's sprawling galaxy distribution was also evident in  
Max Wolf's (1902) catalogue of nebulae in and around Coma
(see Figure 1).
Shapley's (1934) observations led him to conclude that 
``A general inspection of the region within several degrees
of the Coma cluster suggests that the cluster is part of, or
is associated with, an extensive metagalactic cloud...''
Shane \& Wirtanen (1954), on the other hand, argued that
the Coma cluster is a single isolated entity which blends into
the background at a projected distance of $\sim 2^{\circ}$ 
($2.5\,h^{-1}$ Mpc)
from its centre.
Yet even Zwicky (1957), who steadfastly denied the existence of
structure on supercluster scales, nevertheless acknowledged that
the Coma cluster extends to a projected radius of at least 
$6^{\circ}$ from the cluster centre, corresponding to distances
of more than $7\,h^{-1}$ Mpc. 
Abell (1961) examined the distribution of the rich clusters that he
had catalogued and suggested that Coma is one of six members
of a supercluster that extends $\sim 45\,h^{-1}$ Mpc
in diameter.

\begin{figure}[h]
\psfig{figure=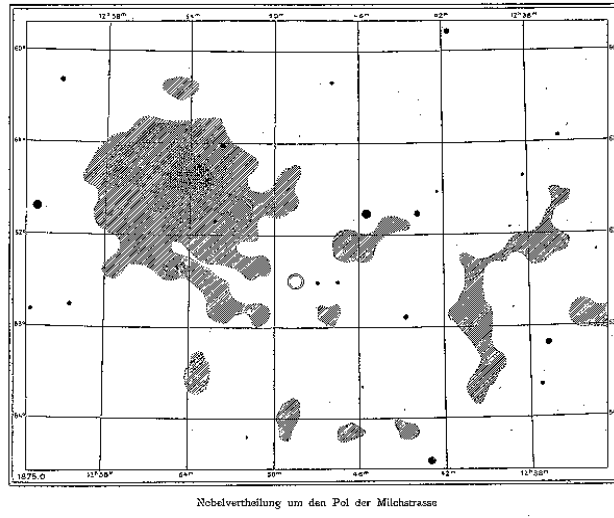,height=4.8in}
\vspace{-2.0in}
\caption{Wolf's (1902) plot of the number density of nebulae in 
and around the Coma cluster. The area shown is $\sim 5^{\circ} \times 7^{\circ}$.
Note the extension to the south-west
which lies $5^{\circ}$ ($\sim 6\,h^{-1}$ Mpc) from the cluster centre.}
\end{figure}

\section{Coma's Cosmic Neighbourhood}

Modern galaxy catalogues and redshift surveys 
have helped bring our view of the 
Coma cluster's environs into much clearer focus.
As Figure 2 shows, Coma is the most conspicuous concentration in
a highly structured large-scale galaxy distribution
that is quite filamentary in appearance, with 
thin quasi-linear features extending over large portions
of the sky.  In particular, a prominent ridge of galaxies
can be seen connecting Coma to the rich clusters A2197, A2199 
and A1367, with several smaller filaments
branching off in different directions.  
A closeup view of a $10^{\circ} \times 10^{\circ}$ degree
region centered on the Coma cluster is shown in Figure 3.
The cluster is clearly embedded in an elongated swath of 
galaxies extending along a projected position angle of
$\sim 70^{\circ} - 80^{\circ}$. 

\begin{figure}
\psfig{figure=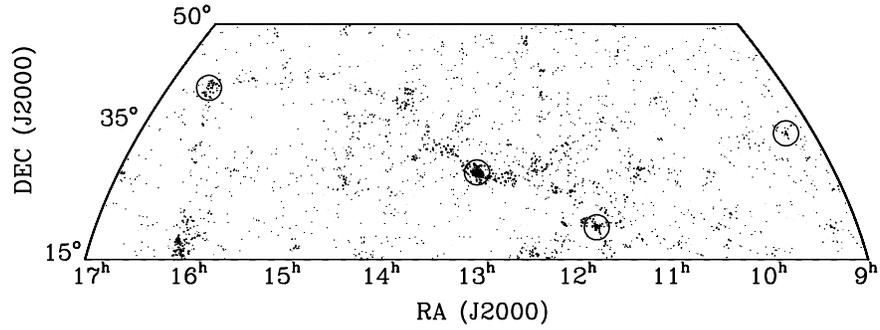,height=4.8in}
\vspace{-3.1in}
\caption{The distribution of 3649 galaxies contained in
the RC3 (de Vaucouleurs et al. 1991).  To highlight features in
the galaxy distribution, symbol sizes are proportional to
local galaxy density (from West 1994). Circles denote Abell clusters
with redshifts $z \leq 0.03$.  In order of increasing right ascension
they are Abell 779, 1367, 1656 (Coma) and 2197/2199.}
\end{figure}

\begin{figure}
\psfig{figure=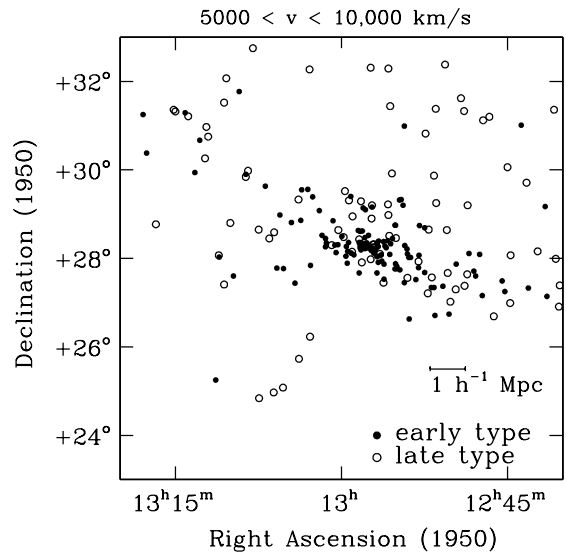,height=3.8in}
\vspace{-1.1in}
\caption{The distribution of Coma galaxies from the catalog of
Doi et al. (1995).
Note the strong spatial segregation between early and late-type galaxies.}
\end{figure}

Gregory \& Thompson's (1978) pioneering redshift survey of 
Coma's environs provided the first direct evidence
that this apparent feature in the projected galaxy distribution
is in fact a genuine three-dimensional bridge linking  
Coma with its nearest neighbouring cluster, Abell 1367, which
lies some $22\,h^{-1}$ Mpc away.
A few years later the 
famed CfA redshift survey (de Lapparent, Geller \& Huchra 1986; 
Geller \& Huchra 1989) mapped the galaxy distribution
surrounding Coma in exquisite detail, producing one of the most
unforgettable astronomical images of the 1980s 
of a majestic $\sim 150\,h^{-1}$ Mpc long
structure sweeping across the entire survey region (see Figure 4).

\begin{figure}[t]
\psfig{figure=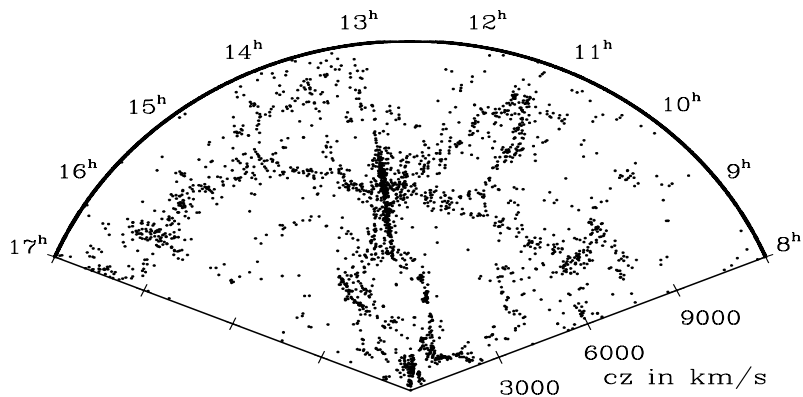,height=4.5in,width=5.5in}
\vspace{-2.3in}
\caption{The distribution of galaxies in velocity-right ascension space
from the CfA redshift survey.
A $12^{\circ}$ slice in declination from 
$26.5^{\circ}$ to $38.5^{\circ}$ is shown here.
The Coma cluster is apparent as the ``finger of god'' elongation
along the line of sight near RA = $13^{h}$. Owing to the 
appearance of the galaxy distribution in this image, it has
often been labelled the ``stickperson'' or the ``homunculus.''}
\end{figure}

\section{Great Walls or Great Filaments?}

Many different adjectives have been used to describe the appearance
of the large-scale galaxy distribution surrounding Coma.
For example,
Tago, Einasto \& Saar (1984) proposed that the Coma-A1367 supercluster
is characterized by a network of interconnected ``cluster chains'' 
or ``filaments.''
de Lapparent, Geller \& Huchra (1986) on the other hand 
suggested that ``the galaxies appear to
be on the surfaces of bubble-like structures'' and argued that  
any filamentary features are merely an artifact
of viewing slices through these bubbles.
Further extensions of the original CfA survey confirmed
that the large-scale feature surrounding the Coma cluster 
remains coherent over
adjacent slices in declination.  Geller \& Huchra (1989)
described this structure as ``sheet-like'' or ``sponge-like'' 
and christened it the ``Great Wall'' noting that it has 
dimensions of at least $150\,h^{-1}$ Mpc $\times$
$60\,h^{-1}$ Mpc but a line-of-sight thickness of only
$\sim 5\,h^{-1}$ Mpc.

Is the large-scale galaxy distribution surrounding the Coma 
cluster filamentary, planar, bubble-like or some combination?
These different descriptions of Coma's environs are 
not merely an exercise in subjective semantics; the topology
of the large-scale matter distribution is expected to be
an important test of different models for the formation of
structure in the universe, and hence an objective and
accurate description is essential.
Geller \& Huchra (1989) argued that ``the thin large-scale features
in this region are cuts through two-dimensional sheets, not one-dimensional filaments'' because if the $\sim 150\,h^{-1}$ Mpc long structure
is a filament then ``we would expect to see corresponding elongated
structures in the distribution projected on the sky'' and none is
observed.  However, Figure 2 shows that strikingly coherent 
filamentary structures are indeed visible when the image is ``enhanced'' to 
bring out the hightest density regions in the projected galaxy distribution.
Similar filamentary features are also seen around other clusters,
most notably the  
Perseus-Pisces supercluster (Haynes \& Giovanelli 1986),
where again a number of smaller filaments appear to merge into
a larger one, like tributaries flowing into a larger river.
Comparison of Figures 2 and 4 suggests that at least
some of the ``sheet-like'' appearance found in the CfA survey
may result from taking broad declination cuts across multiple interconnecting filaments, as the decreased resolution
will inevitably tend to smear out individual thin linear structures.
This is {\it not} to say that genuine wall-like features do not exist;
clearly the space between filaments is not completely empty and the
galaxy distribution appears to remain contiguous over these regions,
however such walls may simply be
the ``webbing'' between more prominent filaments 
(Bond, Kofman \& Pogosyan 1996).  

It is perhaps worthwhile 
to consider the namesake of the Great Wall
of the CfA survey, the Great Wall of China.
The Great Wall of China reached $\sim 6000$ km in length,
$\sim 7$ m in height and roughly $\sim 9$ m in width, for an
approximate axial ratio of 1,000,000:1:1.
For comparison, the most prominent large-scale feature in the
the CfA survey extends
at least $150\,h^{-1}$ Mpc in length, yet is only $\sim 5\,h^{-1}$
Mpc in projected thickness in its densest regions
and $\sim 5\,h^{-1}$ Mpc along the line of sight,
corresponding to an approximate axial ratio 30:1:1.
Hence a plausible argument can be made
that, despite their names, both the Great ``Wall'' of 
China and the Great ``Wall'' of
the Coma cluster are in reality highly
elongated filamentary-type structures!

Further evidence that 
the supercluster environment of Coma is best characterized
by a filamentary
rather than planar structure
is presented in the next section.

\section{The Cluster-Supercluster Connection}

Mounting observational evidence suggests that the
distribution of matter on supercluster scales exerts a profound
influence on the formation of clusters and their member galaxies.

Binggeli (1982) first showed that the major axes of rich clusters 
exhibit a remarkable tendency
to point towards neighbouring clusters,
over separations of $\sim 15\,h^{-1}$ Mpc or more.
This result has
since been confirmed by numerous other studies (e.g., West 1989;
Rhee, van Haarlem \& Katgert 1992; Plionis 1994).
This tendency for clusters to have preferred orientations with
respect to their surroundings is 
readily seen in the Coma cluster, whose major axis
(as determined from either the distribution of member galaxies
or hot intracluster gas)
lies along the same $70^{\circ} - 80^{\circ}$ position angle 
as the prominent filament that surrounds it.

These observed alignments indicate an important connection between
cluster formation and supercluster environment.
West, Jones \& Forman (1995) showed that the distribution of
subclusters $-$ the building blocks from which clusters are
assembled $-$ also reflects the surrounding 
filamentary distribution of matter, and they  
interpreted this as evidence that cluster formation
proceeds via the anisotropic infall of material sheparded  
along these filaments.  
Theoretical arguments suggest that this sort of infall along
preferred directions is likely to be a generic phenomenon of many
models of structure formation
(e.g., Bond 1987; Bond, Kofman \& Pogosyan 1996; West 1994).
In the Coma cluster the
arrangement of its multiple subclusters clearly reflects the orientation
of the surrounding supercluster filament (see Figure 5).
It is worth emphasizing that in general one would {\it not} expect to observe
frequent cluster-cluster alignments if superclusters are predominantly
planar, since in that case clusters would presumably have random 
orientations along the surface of the sheet
and hence neighbouring clusters would appear aligned
only in instances when a sheet is viewed nearly edge on.  
A large-scale matter distribution with a filamentary pattern
imprinted on it 
provides a much more natural explanation for the origin of the 
cluster alignments.

This ``cross-talk'' between clusters and superclusters 
appears to extend to even smaller scales.  It has been known for 
some time that central dominant galaxies in clusters exhibit a
marked tendency to 
share the same major axis orientation as their parent cluster
(Sastry 1968; Binggeli 1982; 
Porter et al. 1991; West 1994), and hence they also tend to ``point''
along surrounding filaments.
In the case of the Coma cluster, the orientations of the 
two central dominant galaxies,
NGC 4889 and NGC 4874, as well as the projected separation vector
between them, all share the same $70^{\circ} - 80^{\circ}$
position angle of the Coma-A1367 filament. 

\begin{figure}[t]
\psfig{figure=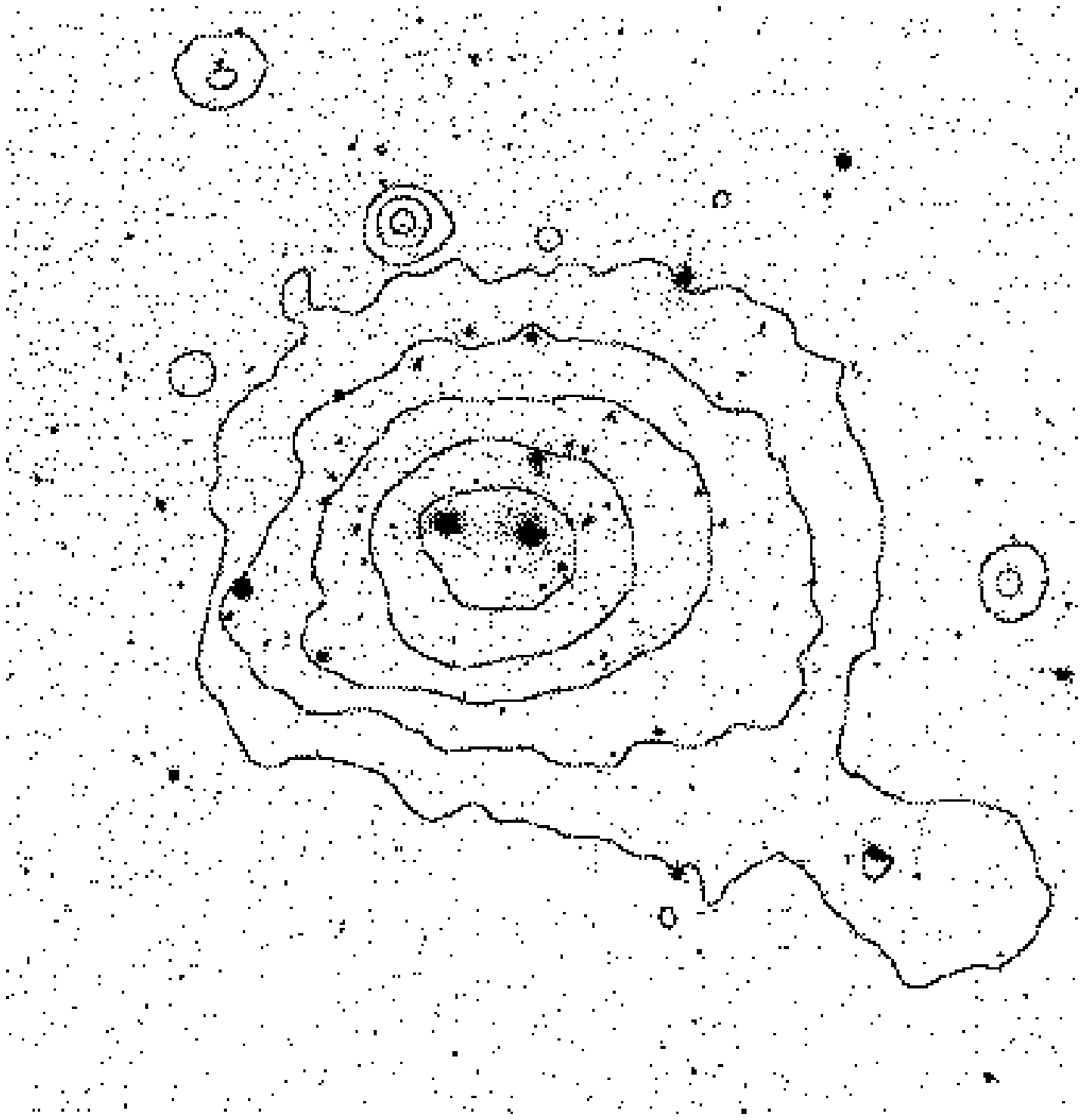,height=2.3in}
\caption{{\it ROSAT} X-ray image of the Coma cluster
(from Vikhlinin, Forman \& Jones 1997).  The subcluster  
associated with the bright galaxy 
NGC 4839 (lower right) appears to be infalling along the same
direction defined by the large-scale filament running from 
Coma to A1367.
Optically identified substructure in the galaxy distribution
in Coma (e.g., Mellier et al. 1988) shows the same effect.}

\vspace{-3.1in}
\psfig{figure=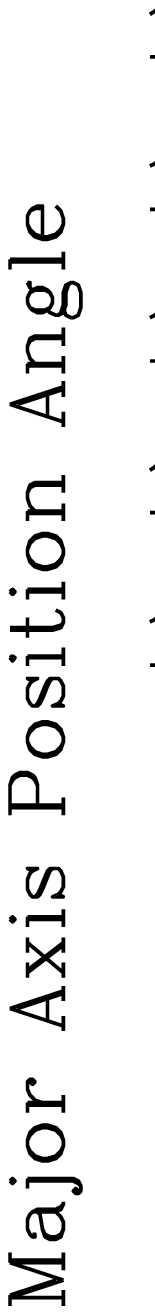,height=2.5in}
\vspace{0.5in}
\caption{The major axis orientation of NGC 4889 as a function
of radius (from West 1994).  Note how the ``preferred'' position
angle can be traced all the way to the very centre of this galaxy.}
\end{figure}

Interestingly, closer examination reveals that this galaxy 
alignment effect is not limited to only the brightest cluster
member.
Table 1 lists the major axis position angles of the brightest
elliptical galaxies in the Coma cluster.  There is a clear tendency for
the orientations of these galaxies to be similar to the 
$70^{\circ} - 80^{\circ}$ position angle defined by the Coma-A1367 supercluster
filament, a fact which was first noted by Brown (1939).
The probability that these galaxy orientations are
random is only $\sim 4\%$ according to a Kolmogorov-Smirnov
test.  Hence the brightest ellipticals in Coma exhibit a statistically
significant tendency to be oriented along the local large-scale
filament.  In fact, in some cases these alignments can be
traced all the way to sub-kpc distances from the galaxy center
(see Figure 6).
The shape and orientation of both the Coma cluster
and its brightest member galaxies clearly reflect the
large-scale anisotropies in the surrounding matter distribution.

A similar effect is seen in the Virgo cluster (Table 1), where again the
brightest elliptical galaxies appear to ``know'' about the
principal axis of the cluster, which in this case lies along a 
projected position angle of $\sim 120^{\circ}$.  
An intriguing possibility is that the Virgo, A1367 and Coma clusters 
are all members of a common filamentary network, with the Virgo
cluster oriented along a filament that runs between it and A1367
(Zeldovich, Einasto \& Shandarin 1982; West 1998).
The tendency for bright elliptical galaxies 
to align with the surrounding
large-scale structure appears to be a rather general phenomenon
(Lambas, Groth \& Peebles 1988;
Muriel \& Lambas 1992).

\begin{table}[t]
\caption{Orientations of the Brightest Elliptical Galaxies in the
Coma and Virgo Clusters}
\vspace{0.2cm}
\begin{tabular}{|l|c|}
\hline
\raisebox{0pt}[13pt][7pt]{Coma Galaxy } &
\raisebox{0pt}[13pt][7pt]{Position Angle} \\
\hline
NGC 4816 & $74^{\circ}$  \cr
NGC 4839 & $61^{\circ}$  \cr
NGC 4841A & $121^{\circ}$  \cr
NGC 4874 & $74^{\circ}$ \cr
NGC 4889 & $77^{\circ}$  \cr
NGC 4911 & $112^{\circ}$  \cr
\hline
\end{tabular}

\vspace{-1.3in}
\noindent\hspace{2.4in}\begin{tabular}{|l|c|}
\hline
\raisebox{0pt}[13pt][7pt]{Virgo Galaxy } &
\raisebox{0pt}[13pt][7pt]{Position Angle} \\
\hline
NGC 4374 (M84) & $135^{\circ}$ \cr
NGC 4406 (M86) & $130^{\circ}$ \cr
NGC 4473 & $100^{\circ}$ \cr
NGC 4478 & $140^{\circ}$ \cr
NGC 4486 (M87) & $110^{\circ}$ \cr
NGC 4564 & $47^{\circ}$ \cr
NGC 4621 (M59) & $165^{\circ}$ \cr
NGC 4649 (M60) & $105^{\circ}$ \cr
NGC 4660 & $100^{\circ}$ \cr
\hline
\end{tabular}
\end{table}

The alignments of galaxies and clusters with their supercluster 
surroundings indicates a 
remarkable coherence of structures over roughly four orders of
magnitude in mass and 
demonstrates clearly that it is the
filaments, and not the walls, that have influenced cluster
formation most strongly.

\section{Conclusions: Lessons to be Learned from Coma}

\begin{itemize}
\item{} Clusters of galaxies are often embedded in large-scale
superclusters which may extend for tens of hundreds of Mpc.

\item{} The morphology of superclusters is primarily (though
not exclusively) filamentary, with quasi-linear bridges of
material connecting neighbouring clusters.

\item{} There can be little doubt that the formation
of the Coma cluster, as well as its brightest member galaxies,
has been driven by infall of material along filaments.
\end{itemize}

What lies in the future for Coma?  Material will
undoubtedly continue to flow into the cluster along filaments, and 
in fact one can even identify future subclusters from the
surrounding distribution of groups (see Figure 7).
Assuming a typical infall velocity of $\sim 1000$ km/s into Coma
(Bothun et al. 1992), Coma is likely to swallow a number of
sizeable groups (Ramella, Pisani \& Geller 1997)
over the course of the next few billion years.

\vspace{1.5in}
\begin{figure}
\psfig{figure=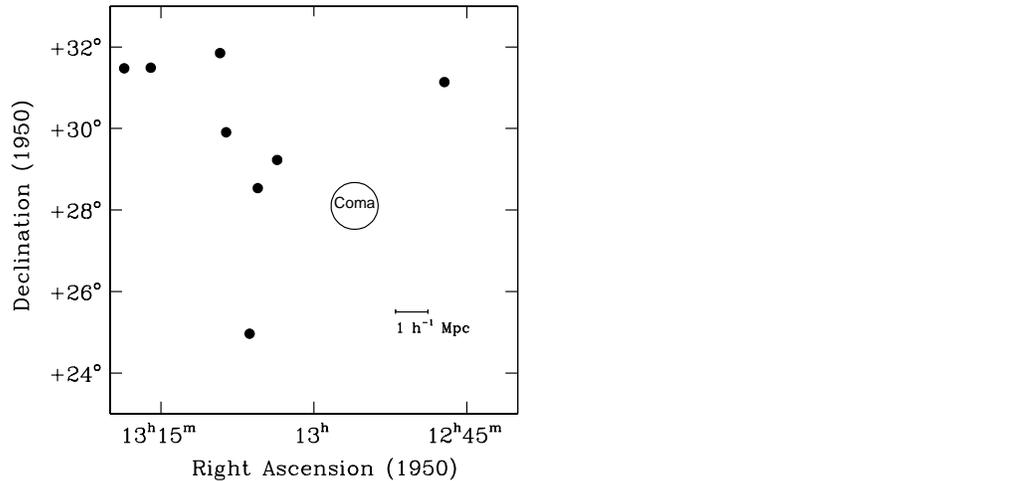,height=3.5in}
\vspace{-0.5in}
\caption{The distribution of groups surrounding the Coma cluster
(from Ramella et al. 1997).  
The majority of these groups, which are the future subclusters
that will fall into the cluster, lie along
the filament which extends to the north-east from Coma.}
\end{figure}

\section*{Acknowledgments}
It is a pleasure to thank Alain Mazure, Florence Durret,
Daniel Gerbal, Fabienne Casoli and everyone on the local 
organizing committee
in Marseille for a most enjoyable and stimulating meeting.
I gratefully acknowledge financial support
from the meeting sponsors, NSERC of Canada and the Canadian
Institute for Theoretical Astrophysics.

\end{document}